\pdfoutput=1

\documentclass[11pt]{article}

\usepackage[]{ACL2023}
\usepackage{float}
\usepackage{times}
\usepackage{latexsym}

\usepackage[T1]{fontenc}

\usepackage[utf8]{inputenc}

\usepackage{microtype}

\usepackage{inconsolata}
\usepackage{graphicx}

\usepackage{listings}
\usepackage{xcolor}

\colorlet{punct}{red!60!black}
\definecolor{background}{HTML}{EEEEEE}
\definecolor{delim}{RGB}{20,105,176}
\colorlet{numb}{magenta!60!black}

\lstdefinelanguage{json}{
    basicstyle=\normalfont\ttfamily,
    numbers=left,
    numberstyle=\scriptsize,
    stepnumber=1,
    numbersep=8pt,
    showstringspaces=false,
    breaklines=true,
    frame=lines,
    backgroundcolor=\color{background},
    literate=
     *{0}{{{\color{numb}0}}}{1}
      {1}{{{\color{numb}1}}}{1}
      {2}{{{\color{numb}2}}}{1}
      {3}{{{\color{numb}3}}}{1}
      {4}{{{\color{numb}4}}}{1}
      {5}{{{\color{numb}5}}}{1}
      {6}{{{\color{numb}6}}}{1}
      {7}{{{\color{numb}7}}}{1}
      {8}{{{\color{numb}8}}}{1}
      {9}{{{\color{numb}9}}}{1}
      {:}{{{\color{punct}{:}}}}{1}
      {,}{{{\color{punct}{,}}}}{1}
      {\{}{{{\color{delim}{\{}}}}{1}
      {\}}{{{\color{delim}{\}}}}}{1}
      {[}{{{\color{delim}{[}}}}{1}
      {]}{{{\color{delim}{]}}}}{1},
}

\title{Decision Knowledge Graphs: Construction of and Usage in Question Answering for Clinical Practice Guidelines}

\author{
  Vasudhan Varma Kandula \\ 
  CSE, IITB / Bombay\\
  \texttt{vasudhanvarma@gmail.com} \And
  Pushpak Bhattacharyya \\ 
  CSE, IITB / Bombay\\
  \texttt{pushpakbh@gmail.com} \\
  }

\begin{document}

\maketitle

\begin{abstract}
In the medical domain, several disease treatment procedures have been documented properly as a set of instructions known as Clinical Practice Guidelines (CPGs). CPGs have been developed over the years on the basis of past treatments, and are updated frequently. A doctor treating a particular patient can use these CPGs to know how past patients with similar conditions were treated successfully and can find the recommended treatment procedure. In this paper, we present a Decision Knowledge Graph (DKG) representation to store CPGs and to perform question-answering on CPGs. CPGs are very complex and no existing representation is suitable to perform question-answering and searching tasks on CPGs. As a result, doctors and practitioners have to manually wade through the guidelines, which is inefficient. Representation of CPGs is challenging mainly due to frequent updates on CPGs and decision-based structure. Our proposed DKG has a decision dimension added to a Knowledge Graph (KG) structure, purported to take care of decision based behavior of CPGs. Using this DKG has shown 40\% increase in accuracy compared to fine-tuned BioBert model in performing question-answering on CPGs. To the best of our knowledge, ours is the first attempt at creating DKGs and using them for representing CPGs. 
\end{abstract}

\section{Introduction}
\label{sec:introduction}
Clinical Practice Guidelines (CPGs) are a set of systematically developed statements intended to assist a doctor or a practitioner to make decisions about appropriate health care to be given to a patient under a specific clinical circumstance. CPGs are built based on evidence from past treatments including the patient’s symptoms, conditions over time, and what decisions led to successful treatment. CPGs can change the process of treatment, and outcome of care, improve the quality of care and enable efficient use of resources. Since CPGs are large documents, a lot of time will be taken to manually search CPGs. There is no existing suitable representation for CPGs to perform tasks like searching, navigating, and question-answering. As a result, doctors and practitioners have to manually refer to the guidelines.

 Our \textbf{motivation} is as follows: According to American Hospital Association (\citet{aha2022us}), in 2022, there were more than 33 million admissions of patients in hospitals in the US, which is an average of 91,000 admissions per day. As the number of patients is increasing, there is heavy workload on doctors, and they may have limited time to review and implement complex guidelines. Also, doctors may be unfamiliar with CPGs due to lack of training, and frequent changes in guidelines over time. Lack of familiarity with CPGs can be a barrier to their use in clinical practice, as doctors may not be aware of the most up-to-date recommendations or may not know how to apply the guidelines to their patients. Therefore, to promote the usage of CPGs, the above barriers need to be overcome. One way to achieve this is by digitizing the guidelines and providing assistance when referring the guidelines using technology.

 The existing Knowledge Graph representation on which searching and question-answering can be performed is not suitable for storing CPGs as CPGs contain a decision-based structure along with factual data and these decisions in CPGs are updated frequently. 
 Given the following guideline:
 \begin{quote}
      \textit{"Patient can be treated with chemotherapy if age less than 65"}
 \end{quote}
 The existing KG extraction model gave:
Subject: \textit{Patient}; Predicate: \textit{can be treated with}; \textit{Object: chemotherapy}. Therefore, the extracted triple is \textit{(patient, can be treated with, chemotherapy)}. The model ignored the condition of age less than 65, which is important for guiding the doctor. Therefore, a good CPG knowledge graph should represent not only concepts but also decisions (attributes). If the above guideline is updated to:
 \begin{quote}
      \textit{"Patient can be treated with chemotherapy if age less than 65 and greater than 35. He should not have any substantial comorbidities."}
 \end{quote}
The existing KG model will require many changes in its structure (i.e, number of nodes and relations). A good CPG knowledge graph representation should have an efficient updating capability with few changes. 

Now-a-days with pretrained models which are performing well in question answering tasks the limitation is that there is no sufficient data for training the model. And even if we create huge data and train the model since the treatment guidelines are changed frequently the dataset should also be updated with the guidelines which is another limitation. Considering this storing the CPGs seems better approach compared to pretraining the models.
 \\
Our contributions are:
\begin{enumerate}
    \item Creation and releasing of a knowledge graph (KG) with an additional decision dimension added to some nodes in existing KG structure for storing clinical practice guidelines, \textit{i.e., Decision Knowledge Graph (DKG)}.
    \item Creation of dataset of triples containing 8300 questions from acute lymphoblastic leukemia, kidney, and bone cancer. Each \textit{triple} consists of \textit{question}, \textit{answer}, and \textit{cypher query} (used to query decision knowledge graph).
    \item Question-answering model on Clinical Practice Guidelines with the help of Decision Knowledge Graphs. The proposed model gives 40\% better results compared to fine-tuned transformer question-answering model.
\end{enumerate}
To the best of our knowledge, ours is the first attempt at (i) creating a knowledge graph for CPGs and (ii) adding a decision dimension to a node in KG.

The rest of the paper is organized as follows. Section\,\ref{sec:relatedwork} presents a brief survey of the literature. Section\,\ref{sec:cpgs} introduces CPGs along with NCCN Guidelines. In section\,\ref{sec:data_creation} provides details about question-answering dataset creation. Section\,\ref{sec:dkg_builder} explains the DKG structure along with the construction and usage of DKG. Section\,\ref{sec:qa} provides an application of DKG i.e., question-answering on CPGs. Section\,\ref{sec:results} provides the results and analysis. Section\,\ref{sec:conclusion} summarizes and concludes the paper. 

\section{Related work}
\label{sec:relatedwork}
CPGs are written based on evidence, aiming to improve the quality and efficiency of medical treatment and care. They are useful to a doctor in providing proper insights when he/she is treating a patient. Many physicians don't use CPGs. \citet{10.1001/jama.282.15.1458} claims that the main reasons for not using CPGs are their complexity, unfamiliarity, and distrust. Trust can be improved once CPGs start gaining positive attention and lead to successful treatment of patients. Complexity and familiarity need to be addressed for the usage of CPGs. CPGs were introduced in the early 90s yet their familiarity is still a problem in the medical domain.

Given the structured nature, and factual data present in the CPGs, it is reasonable to organize this information as a Knowledge Graph. \citet{rossetto2020lifegraph} describes Knowledge Graph (KG) as static graph triples. If the data is static, KG, once constructed, needs no modifications and can be used to perform question-answering and searching tasks. Once the KG is constructed, modifying the KG is costly and takes time as modification involves updating, changing, or deleting multiple nodes and relations which can propagate. Therefore, at times, KG needs to be reconstructed because of some modifications.

Construction of a KG involves many steps like co-reference resolution, information extraction, etc. \citet{kgen} provides a detailed pipeline of KG construction for biomedical scientific literature. Many existing approaches to constructing KG ignore the conditional statements that are present in the sentences. \citet{multi} explains how existing ScienceIE models capture factual data and will not consider conditional statements. i.e., An existing system would return the tuple (alkaline pH, increases, activity of TRPV5/V6 channels in Jurkat T cells) if the statement "alkaline pH increases the activity of TRPV5/V6 channels in Jurkat T cells" was given. However, in this case the condition tuple (TRPV5/V6 channels, in, Jurkat T cells) was not identified.

\citet{bkgc} emphasizes the importance of conditional statements in biomedical data. They also propose a KG representation with conditional statements. The conditional statements are added to the existing KG structure but this structure is not suitable for clinical practice guidelines because updating is not efficient in the current KG structure.

From the survey conducted by \citet{liang2022reasoning}, many KG question-answering models were relying on rules, keywords, neural networks, etc. After the introduction of SPARQL by \citet{hu2021natural}, which is a query language to search and modify a KG, retrieving data from KG became easy. Therefore, many question-answering models were proposed using KG.

The existing representations of CPGs are complex and unfamiliar as mentioned in \citet{10.1001/jama.282.15.1458}. Manually searching data in CPGs takes time. During emergencies, time is valuable and lack of time can cost lives. A representation for CPGs on which question-answering and searching can be performed will help a lot in emergencies. This representation can also motivate practitioners and doctors to use guidelines. So far, no attempt has been made for representing CPGs to perform question-answering and searching tasks.
\section{Background}
In this section, we briefly describe decision knowledge graph and question-answering system. 
\subsection{Decision Knowledge Graph}
\label{subsec:dkg}
Decision knowledge graph is a knowledge graph structure with decision dimension added to its structure. We store data related to patients' parameters and conditions of patient in decision dimension. This data is called as \textit{Patient's Constraints} which are often referred to as \textit{Constraints} in rest of the paper. Some of the examples of patients' constraints are \textit{Age, tumor size, disease stage, past medical history, etc.} We divide data into static and dynamic data. Static data refers to the data in Clinical Practice Guidelines (CPGs) which changes less frequently or doesn't change at all. \textit{Example: Treatment procedure like chemotherapy etc.} Dynamic data refers to the data in the CPGs which changes frequently. Here, dynamic data doesn't refer to data from a query like the name of the patient, etc. It refers to the data that should be present in the KG to make a decision. \textit{Example: Patient constraints}.
\subsection{Question-Answering System}
A question-answering system is a model which is trained to generate correct answer to given question. There are many ways to approach question-answering. One of the ways is language model trained on input-output pairs such that input is a question and output is the answer.

\section{Clinical Practice Guidelines for Cancer}
\label{sec:cpgs}
\begin{figure}[h]
    \centering
    \includegraphics[width=8cm]{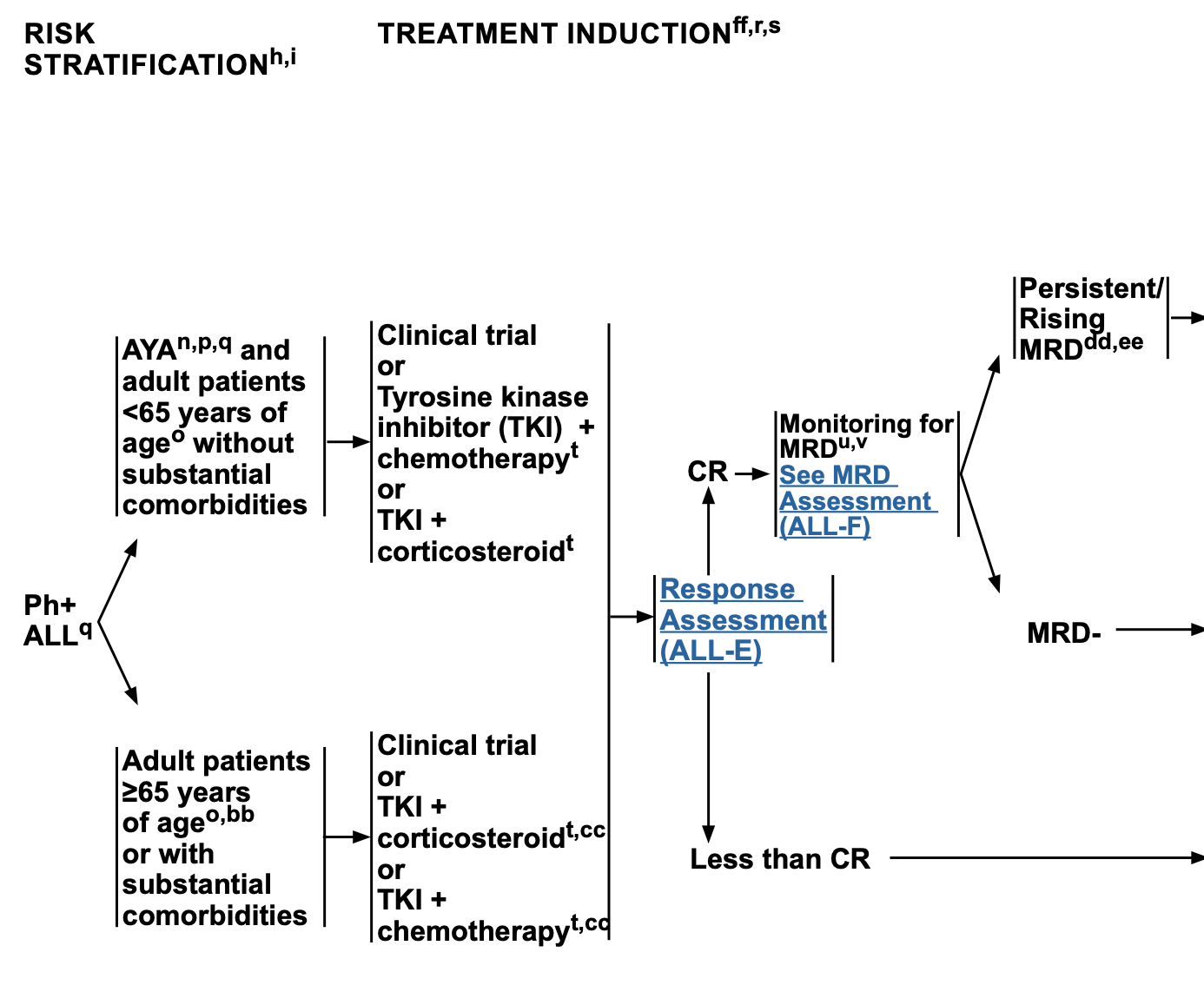}
    \caption{Fragment of Clinical Practice Guidelines by National Comprehensive Cancer Network from page 12 of Acute Lymphoblastic Leukemia (ALL) cancer Version 1.2022 which shows how a ph+ (Philadelphia chromosome) ALL patient should be treated in the induction phase of ALL cancer. Refer to Appendix \ref{apx:cpgexample} for detailed explanation of above fragment}
    \label{fig:CPGFragmentFig}
\end{figure}

Clinical Practice Guidelines (CPGs) from National Comprehensive Cancer Network (NCCN) are used for building Decision Knowledge Graph (DKG). These are also referred to as Cancer Guidelines, NCCN Guidelines, or Oncology Guidelines. NCCN is a non-profit alliance dedicated to facilitating effective, quality, and accessible cancer care. The organization is home to around 60 types of cancer research and guidelines including breast cancer, lung cancer, kidney cancer, etc. For the past 25 years, these guidelines are updated regularly based on discussions among world-renowned experts from NCCN member institutions. A snapshot of the NCCN Guidelines, taken from page 12 of Acute Lymphoblastic Leukemia (ALL) Cancer Version 1.2022, is shown in Figure\,\ref{fig:CPGFragmentFig}.

The NCCN guidelines include:
\begin{enumerate}
    \item List of members and institutions that participated in the specified discussions.
    \item Flowcharts for better understanding of decision making. 
    \item Discussions to provide support for flowcharts.
    \item Evidence for recommendations and disclosure of potential conflicts of interest by panel members (members who attended the discussion).
\end{enumerate}
 The flowchart section of guidelines consists of text boxes and arrows connecting these boxes as shown in Figure\,\ref{fig:CPGFragmentFig}. Some of the words in the text have superscripts and subscripts. Superscripts and subscripts contain a detailed description in the footnote of the paper. There are hyper-texts in some text that refer to other pages in the same document. For more details on CPGs used for this paper refer to Appendix \ref{apx:guidelines}.

\section{Dataset Creation}
\label{sec:data_creation}
The main objective of a Decision Knowledge Graph (DKG) is to perform question-answering thus reducing the manual effort of a doctor to search through the guidelines. There are no available question-answering datasets on Clinical Practice Guidelines. We have created a CPG-QA dataset with 8300 question-answer pairs. This dataset consists of three main types of questions. 
Types of questions:
\begin{enumerate}
    \item \textbf{What is next treatment advice given a patient's constraints (refer to Section \ref{subsec:dkg} for more details on constraints).}\\
    \textit{Example: A patient is ALL positive. After his initial diagnosis he is classified as ph- patient. His age is 65. He is not treated with other cancer treatments. What treatment is recommended in this condition?}
    \item \textbf{What are the patient's medical constraints that needs to be satisfied given a treatment stage.}\\
    \textit{Ex: A patient is ALL positive. After his initial diagnosis he is classified as ph+ patient. What are patient constraints for doing chemotherapy?}
    \item \textbf{Given a patient's medical constraints and treatment stage, whether a particular treatment is advisable or not?}\\
    \textit{Ex: A patient is ALL positive. After his initial diagnosis he is classified as ph- patient. His age is 65. He is not diagnosed with any other cancer treatment. Can we perform TKI + Chemotherapy on him?}
\end{enumerate}
The dataset also consists of cypher queries for question-answering pairs which are used to query the DKG. These cypher queries are manually constructed given a question. We have verified the correctness of the queries by running them on DKG and matching the outputs of DKG with the expected answer. The format of the dataset is:
\begin{lstlisting}[language=json,firstnumber=2]
[
    {
      "QUESTION": String,
      "ANSWER": String,
      "QUERY": String,
      "Expected_Node": Integer,
      "DKG_response": Integer,
    },...
]

\end{lstlisting}
Examples can be referred from Appendix \ref{apx:dataset}.

\section{Decision Knowledge Graphs}
\label{sec:dkg_builder}
This section presents the decision knowledge graph (DKG), its construction, and details on how operations like updating, deleting, and insertion, can be performed on DKGs. 
\subsection{Introduction}
In the Knowledge Graph (KG), data is stored as triples consisting of a head entity, a relation, and a tail entity i.e., (head, relation, tail). If there is some change in the KG (i.e., updating triple, deleting triple, or adding new triple), these changes, in the worst case, can propagate to all nodes. Consider the example \textit{ given triple (Barack Obama, president of , US) if we want to update Obama to Trump then the update should be done in multiple nodes which talk about US presidency or about the individuals}. Therefore sometimes, updating a KG will become equivalent to rebuilding the KG. The update operation, therefore, is time-consuming. Clinical Practice Guidelines (CPGs) are updated frequently. Hence, KG structure won't be of much help for CPGs as it would require the costly update operation frequently. 

From the previous few versions of guidelines, we have observed that not all content in the guidelines is changed. The modifications that are made to guidelines, based on discussions, are mainly done on patients' constraints (refer to Section \ref{subsec:dkg} for definition). The treatment steps of chemotherapy are not changed but when to perform chemotherapy based on the patient's condition is changed. Therefore, using this observation, we divide the data into static and dynamic data. 

Static data is the data in CPGs that changes less frequently or doesn't change at all. Dynamic data is the data in CPGs which changes frequently. Here, dynamic data doesn't refer to data from a query like the name of the patient, etc. It refers to the data that should be present in the KG to make a decision. \textit{For example, treatment procedure like chemotherapy is static data and patients’ constraints like \textit{age>60, MRD rising,} etc., is dynamic data. }

DKG is a knowledge graph over which we have introduced a decision layer. This decision dimension will consist of dynamic data. Static data is stored as KG triples extracted as proposed by \citet{kgen}. For example, if there is a node, "chemotherapy", we have relations like "procedure", "drugs used", "duration" etc., which comes under static data.  When updating a KG, only dynamic data needs to be changed without changing the structure of the KG and static data. Therefore, performing updates on DKG will be a more cost-effective task than updating a KG. Here the static data is stored as a KG. For example, if there is a node, “chemotherapy”, we have relations like “duration”, “drugs used” etc. Therefore, factual data is stored as we do in a KG, but conditional data is stored in decision nodes.

\begin{figure*}[ht]
    \centering
    \includegraphics[width=17cm]{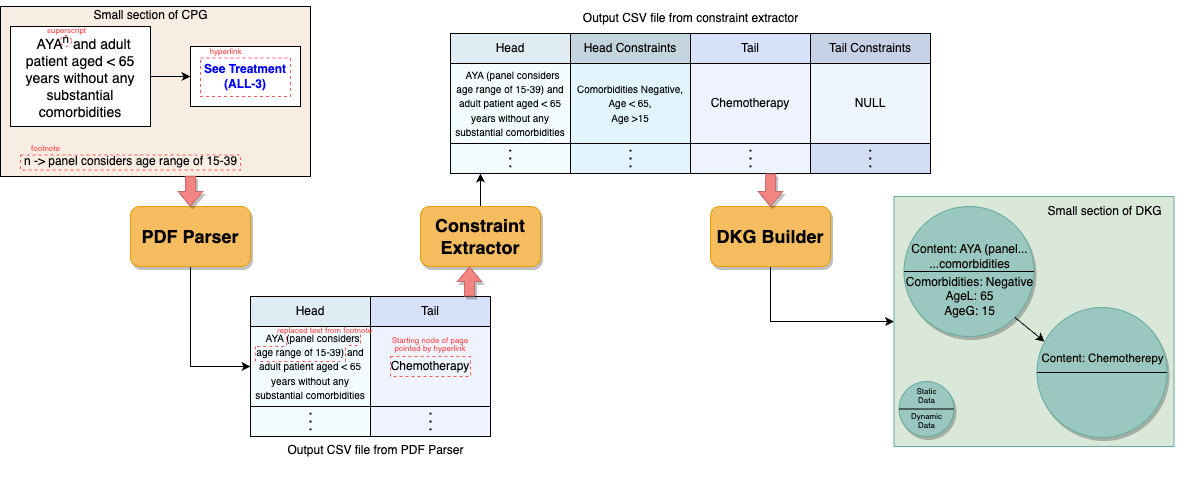}
    \caption{DKG Construction; i) PDF Parser: converts PDF of NCCN guidelines to CSV file, ii) Constraint Extractor: extracts the constraints (refer to Section \ref{subsec:dkg} for definition) from each sentence and adds them to CSV file, iii) DKG Builder: takes the CSV and builds the DKG in neo4j graph database}
    \label{fig:arch}
\end{figure*}

\subsection{Construction of Decision Knowledge Graph}
DKG is constructed by three main modules as shown in Figure \ref{fig:arch}: PDF Parser, Constraint Extractor, and DKG builder.

\subsubsection{PDF Parser}
\label{sec:pdf_parser}
Input to the PDF parser is the CPG PDF file. The PDF Parser recognizes the text in text boxes in the CPGs using optical character recognition (OCR). Superscripts and subscripts on text, as described in Section \ref{sec:cpgs}, are replaced with the text given in the footnotes. Hypertexts, described in Section \ref{sec:cpgs}, in the text boxes, are replaced with the content that it is pointing to. The output of the PDF parser is a CSV file with two columns: the first column corresponds to the head entity (text present in the box of the arrow tail), and the second column corresponds to the tail entity (text present in the box of the arrowhead).  

\subsubsection{Constraint Extraction}
\label{sec:ce}
The constraint extractor iterates over each sentence in the CSV file generated above. On each input sentence, it outputs the constraints (refer to Section \ref{subsec:dkg} for definition) in the sentence. If there are no constraints in a sentence, NULL is returned. If there are multiple constraints, they are returned separated by a comma (,).  

The Constraint extractor is a hybrid (rule-based and deep learning-based) model which uses the output of a constituency parser. In constraint extractor, the input sentence is first pre-processed, and the pre-processed sentence is tokenized and passed to the constituency parser. The output of the constituency parser is a tree-based structure (refer Appendix \ref{apx:consti} for more details). The tree nodes are merged recursively with regular expression rules for linking the entities which are close to each other. Stop words and verbs are removed from the sentence and mathematical words are replaced by their symbol. This final output is given to a keyword-based extractor to get constraints.

The output of the constraint extractor is stored in the constraint column in the CSV file along with the sentence. 

\subsubsection{DKG Builder}
The above generated CSV file has four columns: Head entity, Head Constraints, Tail entity, and Tail Constraints. These will be used to build the DKG. The head entity is a sentence, present as data in the head node and head constraints are the patients’ constraints, separated by a comma (,). Similarly, tail entity and tail constraints are tail node data and patients’ constraints. The head entity and the tail entity will be stored as static data, and the head and tail constraints as dynamic data.  We have used the neo4j graph database (licensed and distributed under GPL v3) to store this knowledge graph. Loading the CSV file to neo4j can be done using \textit{“LOAD CSV FROM <path\_to\_csv>”} command. As the neo4j graph database allows multiple property-value pairs in a single node, we have stored static data with property name “content” and constraints with property name depending on the type of constraint as shown in Figure \ref{fig:arch}.
\begin{figure*}[h]
    \centering
    \includegraphics[width=16cm]{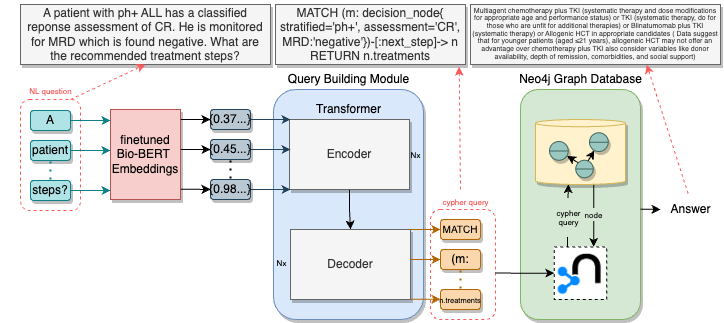}
    \caption{Question Answering using DKG; i) Query Building Module builds the cypher query from given natural language (NL) question, ii) Neo4j Graph Database fetches the node from the DKG according to the query and returns the content of the node}
    \label{fig:qa_dkg}
\end{figure*}
\subsection{Searching in Decision Knowledge Graph}
\label{subsec:cql}
We have used Cypher Query Language (CQL) to query DKG. CQL is like Structured Query Language (SQL). SQL is used to query famous database management systems like PostgreSQL, MySQL, etc., while CQL is used to query the neo4j graph database. 

The syntax used by CQL is of the ASCII-art variety, with \textit{(nodes)-[: ARE\_CONNECTED\_TO]->(otherNodes)} employing rounded brackets for circular \textit{(nodes)} and \textit{-[: ARROWS]->} for relationships. It creates a graph pattern over the data when we write a query. We can use \textit{MATCH} query to search the DKG. If we want to know the next treatment step for a patient who is \textit{ph+ ALL} and Minimal Residual Disease (MRD) is \textit{rising}, then the corresponding CQL query will be: \textit{MATCH (m: node{stratified=`ph+', MRD:`rising'})-[:next\_step]-> n RETURN n.treatments}. Here, \textit{m} and \textit{n} are node variables.

\subsection{Operations on Decision Knowledge Graph}
We can perform the following operations on a DKG: deleting a constraint, inserting a new constraint, and updating a constraint. Deleting a constraint can be done using the command \textit{``MATCH node REMOVE constraint''}. Inserting a constraint can be done using the command \textit{``MATCH node SET constraint''}. Updating can be done by deletion followed by insertion. The time taken for performing the above operations is search time taken by \textit{MATCH} operation, which is \textit{O(nodes)} (linear), as \textit{SET} and \textit{REMOVE} operation takes \textit{O(1)} (constant) time. 

\subsection{Constructed DKG Information}
The DKG is generated for three types of cancers, ALL, Bone, and Kidney. Table \ref{tab:dkgs} shows the information on the number of nodes and relations in these DKGs.
\begin{center}
\begin{table}[ht]
\begin{tabular}{ |c|c|c|c| } 
\hline
\textbf{Cancer type} & \textbf{Total} & \textbf{Decision} & \textbf{Relations} \\
\hline
\textbf{ALL} & 58 & 20 & 74 \\ 
\textbf{Bone} & 191 & 72 & 243 \\ 
\textbf{Kidney} & 50 & 16 & 61 \\
\textbf{Total} & 299 & 108 & 378 \\
\hline
\end{tabular}
\caption{\label{tab:dkgs}Results showing number of nodes and relations in DKG. $1^{st}$ col specifies the cancer type, $2^{nd}$ col specifies total number of nodes in the DKG structure, $3^{rd}$ col specifies total number of decision nodes, and $4^{th}$ col specifies total number of relations in the DKG structure. } 
\end{table}
\end{center}


\section{Question-Answering on Clinical Practice Guidelines (CPGs)}
\label{sec:qa}
In this section, we discuss the models used to perform question-answering.  

\subsection{Word Embeddings}
\begin{figure}[ht]
    \centering
    \includegraphics[width=8cm]{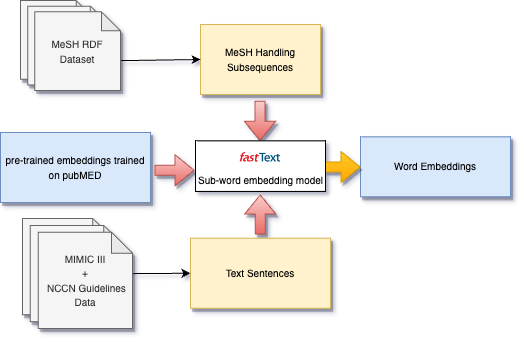}
    \caption{Model to generate embeddings for missing words and improve existing embeddings from NCCN guidelines}
    \label{fig:embeddings}
\end{figure}

BioBERT from \citet{lee2020biobert} is a pre-trained biological language representation model based on the BERT from \citet{devlin2018bert} (Bidirectional Encoder Representations from Transformers) architecture, which is a natural language processing neural network model. BioBert is pre-trained on a huge corpus of biomedical texts, such as PubMed, making it especially well-suited for biomedical text mining and related applications. It is pre-trained to capture the nuances of biomedical language and terminology, and has shown state-of-the-art performance on various biomedical tasks. As BioBERT has missing embeddings for words used in NCCN guidelines. We have created new embeddings using the architecture shown in Figure \ref{fig:embeddings}. We have used MeSH RDF dataset for domain knowledge i.e., we have checked whether the subword from NCCN guidelines is present in MeSH data or not. If the subword is not present, we have avoided training with the particular subword. Datasets of NCCN guidelines and MIMIC III are augmented for training. Subword embedding model from fasttext (MIT License) is used for training. Embedding correctness is checked using analogy task. 
\subsection{Question-Answering without DKG}
\begin{figure}[ht]
    \centering
    \includegraphics[width=8cm]{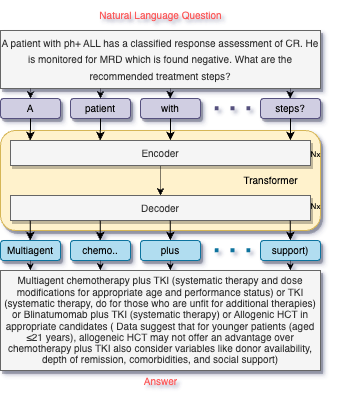}
    \caption{Question Answering without DKG; transformer model trained on question and answers from the guidelines}
    \label{fig:qa}
\end{figure}

Figure \ref{fig:qa} shows the architecture of the model. A transformer is used to perform question-answering (QA) task. Here, the model takes a question (natural language question specifying the conditions of the patient) and generates an answer (recommended next treatment procedure). We split the data into 70\% train, 15\% validation and 15\% testing. The model consists of 19 million parameters with 8 heads, 256 latent dimension.
\subsection{Question-Answering with DKG}
Figure \ref{fig:qa_dkg} shows the architecture of the proposed model. As we have seen in Section \ref{subsec:cql}, we need CQL to query DKG. Given a natural language question from the user, using a transformer model, we convert the question to CQL query. We have used the dataset that is created in Section \ref{sec:data_creation} to train the model.  We have post-processed the generated query based on the syntax of CQL. The post-processed query's parameters are verified from the question. This generated CQL query is used to retrieve data from the neo4j database. Neo4j database retrieves the matched node corresponding to the CQL query from the DKG which is the answer to the natural language question. We split the data into 70\% train, 15\% validation and 15\% testing. The model consists of 19 million parameters with 8 heads, 256 latent dimension.

\section{Results and Analysis}
\label{sec:results}

Table \ref{tab:results} shows the results on both question-answering models, with and without DKG. Having DKG has improved accuracy (calculated as number of correct matches divided by total number of questions) by 40\% compared to the deep learning model. The model with DKG has outperformed in every metric. This shows that having the knowledge of guidelines will help in getting better results. The model with DKG is performing better compared to the model without DKG. Some of the reasons of this improvement is dataset size as transformer is data hungry we need a large amount of data to make transformer perform well, and unavailability of domain knowledge in the model without DKG. 

\begin{itemize}
    \item \textbf{Question: } \textit{A 68-year-old ph-ALL patient without any significant comorbidities underwent a clinical trial during the treatment induction phase, achieving a CR response assessment. He was monitered with persistent rising MRD. What procedures are recommended?}
    \item \textbf{Actual Answer: } \textit{Blinatumomab follwed by Allogenic HCT}
    \item \textbf{Predicted Answer (without DKG): } \textit{Predicted Answer: Allogenic HCT (especially if high-risk features or consider continuing multiagent chemotherapy or Blinatumomab}
    \item \textbf{Predicted cypher query: } \textit{MATCH (m: decision\_node{ stratified='ph-', MRD:'rising'})-[:next\_step]-> n RETURN n.treatments}
    \item \textbf{Predicted Answer (with DKG): } \textit{Blinatumomab follwed by Allogenic HCT}
    
\end{itemize}
\begin{center}
\begin{table}[ht]
\begin{tabular}{ |c|c|c| } 
\hline
\textbf{Metric} & \textbf{Without DKG} & \textbf{With DKG} \\
\hline
ROUGE precision & 0.49 & 0.95 \\ 
ROUGE recall & 0.62 & 0.96 \\ 
ROUGE f-measure & 0.51 & 0.96 \\ 
BLEU & 0.44 & 0.95 \\
Jaccard & 0.46 & 0.92 \\
Accuracy & 0.259 & \textbf{0.676} \\
\hline
\end{tabular}
\caption{\label{tab:results}Results on QA with DKG and without DKG; $1^{st}$ col corresponds to various metrics; the baseline model (the $2^{nd}$ col) is a fine-tuned Bio-Bert model as described in Figure \ref{fig:qa}; the proposed model (the $3^{rd}$ col) is a transformer model with Decision Knowledge Graph (DKG) support as described in Figure \ref{fig:qa_dkg}, Metric definitions can be referred from Appendix \ref{apx:evm} }
\end{table}
\end{center}

\section{Conclusion and Future Work}
\label{sec:conclusion}

In conclusion, representing clinical practice guidelines (CPGs) digitally is challenging. The proposed novel structure, Decision Knowledge Graph (DKG) can effectively store CPGs. DKG enables the encoding of decision-based structures, which are often changed in CPGs, in addition to factual data. Our work makes a significant addition to the field of representing medical knowledge and can help practitioners and doctors to make well-informed judgments about patient's treatment. Our work also contributes to the NLP community by providing a representation for storage of knowledge which has decision-based structure. The model is intended to be used by professional practitioners and doctors only and for recommendation purpose, not to solely depend on the models recommended treatment. 

The DKG is constructed only for NCCN guidelines in this paper. But this structure can be used for other guidelines data. The structure is not restricted to medical domain but can also be expanded to other domains like construction guidelines in Civil engineering, etc.


\section*{Limitations}
The model can suggest recommended treatment procedures for ALL cancer type based on NCCN guidelines version 1.2022 of ALL cancer. This recommended treatment still needs the involvement of doctor. It does not replace the work done by doctor, instead helps him in making things faster. The work done is limited to CPGs, and data having decision based behaviour. DKG is not useful to store he data which don't have this behavior.

\bibliography{anthology,custom}

\begin{thebibliography}{10}
\expandafter\ifx\csname natexlab\endcsname\relax\def\natexlab#1{#1}\fi

\bibitem[{aha()}]{aha2022us}

\newblock Fast facts on u.s. hospitals, 2022.
\newblock \url{http://https://www.aha.org/statistics/fast-facts-us-hospitals}.
\newblock Accessed: 2023-02-15.

\bibitem[{Cabana et~al.(1999)Cabana, Rand, Powe, Wu, Wilson, Abboud, and
  Rubin}]{10.1001/jama.282.15.1458}
Michael~D. Cabana, Cynthia~S. Rand, Neil~R. Powe, Albert~W. Wu, Modena~H.
  Wilson, Paul-André~C. Abboud, and Haya~R. Rubin. 1999.
\newblock \href {https://doi.org/10.1001/jama.282.15.1458} {{Why Don't
  Physicians Follow Clinical Practice Guidelines?A Framework for Improvement}}.
\newblock \emph{JAMA}, 282(15):1458--1465.

\bibitem[{Devlin et~al.(2018)Devlin, Chang, Lee, and
  Toutanova}]{devlin2018bert}
Jacob Devlin, Ming-Wei Chang, Kenton Lee, and Kristina Toutanova. 2018.
\newblock Bert: Pre-training of deep bidirectional transformers for language
  understanding.
\newblock \emph{arXiv preprint arXiv:1810.04805}.

\bibitem[{Hu et~al.(2021)Hu, Duan, and Dang}]{hu2021natural}
Xin Hu, Jiangli Duan, and Depeng Dang. 2021.
\newblock Natural language question answering over knowledge graph: the
  marriage of sparql query and keyword search.
\newblock \emph{Knowledge and Information Systems}, 63(4):819--844.

\bibitem[{Jiang et~al.(2020)Jiang, Zeng, Zhao, Qin, Liu, Chawla, and
  Jiang}]{bkgc}
Tianwen Jiang, Qingkai Zeng, Tong Zhao, Bing Qin, Ting Liu, Nitesh~V Chawla,
  and Meng Jiang. 2020.
\newblock Biomedical knowledge graphs construction from conditional statements.
\newblock \emph{IEEE/ACM transactions on computational biology and
  bioinformatics}, 18(3):823--835.

\bibitem[{Jiang et~al.(2019)Jiang, Zhao, Qin, Liu, Chawla, and Jiang}]{multi}
Tianwen Jiang, Tong Zhao, Bing Qin, Ting Liu, Nitesh Chawla, and Meng Jiang.
  2019.
\newblock Multi-input multi-output sequence labeling for joint extraction of
  fact and condition tuples from scientific text.
\newblock In \emph{Proceedings of the 2019 Conference on Empirical Methods in
  Natural Language Processing and the 9th International Joint Conference on
  Natural Language Processing (EMNLP-IJCNLP)}.

\bibitem[{Lee et~al.(2020)Lee, Yoon, Kim, Kim, Kim, So, and
  Kang}]{lee2020biobert}
Jinhyuk Lee, Wonjin Yoon, Sungdong Kim, Donghyeon Kim, Sunkyu Kim, Chan~Ho So,
  and Jaewoo Kang. 2020.
\newblock Biobert: a pre-trained biomedical language representation model for
  biomedical text mining.
\newblock \emph{Bioinformatics}, 36(4):1234--1240.

\bibitem[{Liang et~al.(2022)Liang, Meng, Liu, Liu, Tu, Wang, Zhou, Liu, and
  Sun}]{liang2022reasoning}
Ke~Liang, Lingyuan Meng, Meng Liu, Yue Liu, Wenxuan Tu, Siwei Wang, Sihang
  Zhou, Xinwang Liu, and Fuchun Sun. 2022.
\newblock Reasoning over different types of knowledge graphs: Static, temporal
  and multi-modal.
\newblock \emph{arXiv preprint arXiv:2212.05767}.

\bibitem[{Rossanez et~al.(2020)Rossanez, Dos~Reis, Torres, and
  de~Ribaupierre}]{kgen}
Anderson Rossanez, Julio~Cesar Dos~Reis, Ricardo da~Silva Torres, and
  H{\'e}l{\`e}ne de~Ribaupierre. 2020.
\newblock Kgen: a knowledge graph generator from biomedical scientific
  literature.
\newblock \emph{BMC medical informatics and decision making}, 20(4):1--24.

\bibitem[{Rossetto et~al.(2020)Rossetto, Baumgartner, Ashena, Ruosch,
  Pernischov{\'a}, and Bernstein}]{rossetto2020lifegraph}
Luca Rossetto, Matthias Baumgartner, Narges Ashena, Florian Ruosch, Romana
  Pernischov{\'a}, and Abraham Bernstein. 2020.
\newblock Lifegraph: a knowledge graph for lifelogs.
\newblock In \emph{Proceedings of the Third Annual Workshop on Lifelog Search
  Challenge}, pages 13--17.

\end{thebibliography}
\bibliographystyle{acl_natbib}

\appendix
    \section{Constituency Parser}
\label{apx:consti}
A constituency parser as referred in \ref{sec:ce} breaks down a phrase into its constituent elements, which are generally represented by a tree diagram. Each node in the tree represents a component, which might be a single word or a phrase or sentence made up of several words. The constituency parser contributes to the resolution of syntactic ambiguity in natural language phrases. Syntactic ambiguity arises when a statement may be interpreted in several ways, resulting in alternative interpretations and meanings. Consider the line \textit{"without comorbidities of diabetes and liver"}. This statement might be paraphrased as \textit{"without comorbidities of diabetes, liver"} or \textit{"without comorbidities of diabetes and without comorbidities of liver"}. The constituency parser can identify and disambiguate the sentence's constituent elements, resulting in a single, well-formed parse tree that captures the sentence's intended meaning. This aids in ensuring that the right sentence interpretation is employed.

\begin{figure}[h]
    \centering
    \includegraphics[width=8cm]{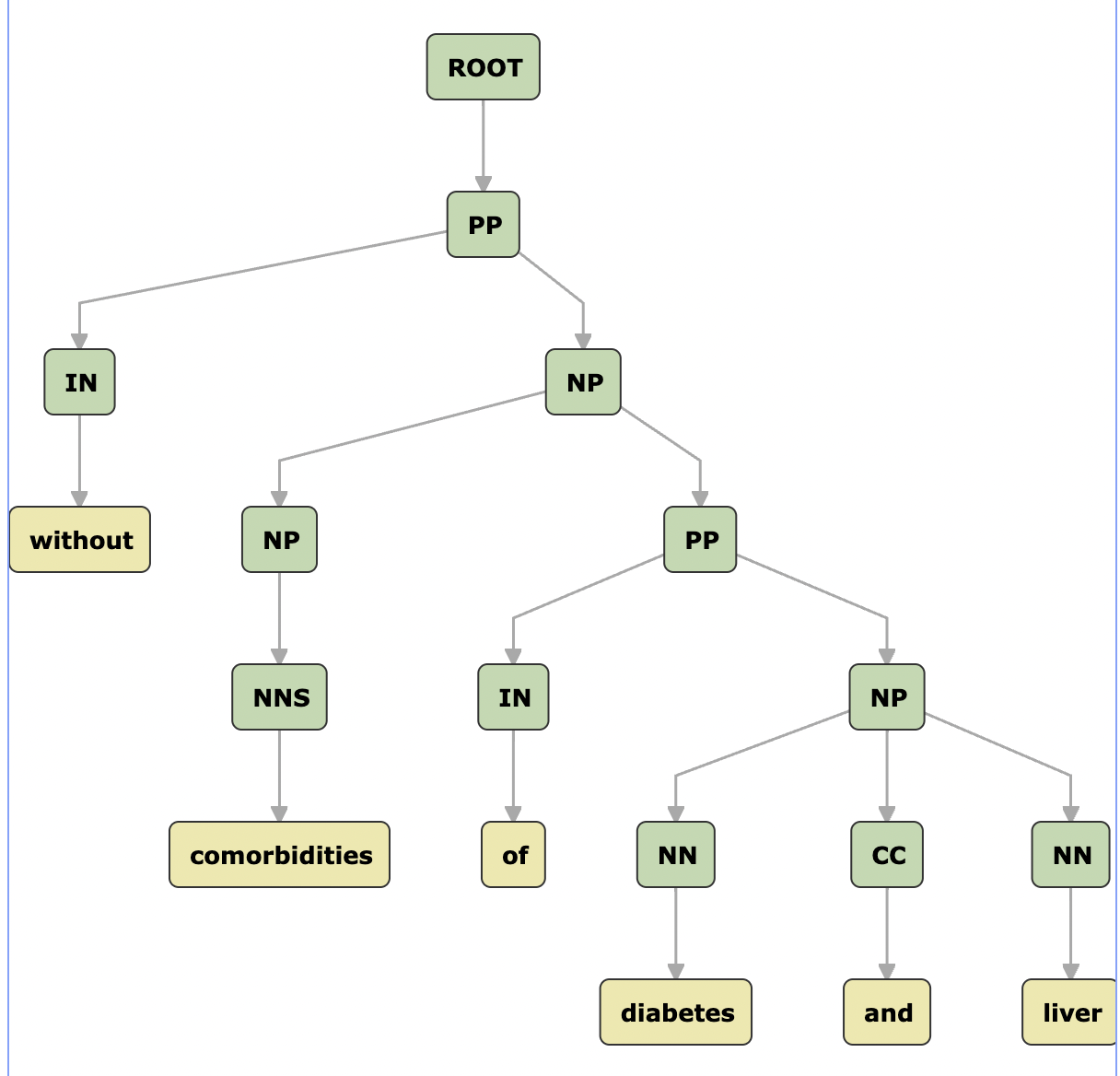}
    \caption{Constituency parser output for sentence \textit{"without comorbidities of diabetes and liver"}, generated using stanford core NLP}
    \label{fig:consti}
\end{figure}

Constituency parser from Stanford CoreNLP is used in Constraint Extractor from Section\,\ref{sec:ce}. Sample output for constituency parser is \textit{Sample output is: (ROOT (S (S (NP (JJ adult) (NNS patients)) (VP (MD should) (VP (VB be) (NP (NP (QP (JJR less) (IN than) (CD 65)) (NNS years)) (PP (IN of) (NP (NN age))))))) (CC and) (PP (IN without) (NP (JJ substantial) (NNS comorbidities))) (. .)))}.

\section{Dataset Examples}
\label{apx:dataset}
Referred in Section \ref{sec:data_creation}.
\begin{lstlisting}[language=json,firstnumber=2]
1. 
{
      "QUESTION": "Upon risk stratification, a patient is identified to have ph- ALL at the age of 37. What treatment measures are advised?",
      "ANSWER": "clinical trial or Pediatric-inspired regimes or Multiagent chemotherapy(systematic therapy)",
      "REMARK": "pediatric-inspired regimes is preferred more",
      "QUERY": "MATCH (n: risk_stratification) WHERE n.stratified = 'ph-' and n.age_cat='AYA' -[:next_step]->k RETURN k.treatment",
      "Expected_Node": 14,
      "DKG_response": 14
}

2. 
{
      "QUESTION": "A ph- ALL patient's response assessment is CR. His age is 37. He was monitored for MRD and found negative. What are the recommended procedures?",
      "ANSWER": "Allogenic HCT (especially if high-risk features or consider continuing multiagent chemotherapy or Blinatumomab",
      "QUERY": "MATCH (m: decision_node{ stratified='ph-', age_cat='AYA', MRD:'absent'})-[:next_step]-> n RETURN n.treatments",
      "Expected_Node": 17,
      "DKG_response": 17
}
\end{lstlisting}




\section{Evaluation Metrics}
\label{apx:evm}
We briefly describe the metrics used in the evaluation reported in Section \ref{sec:results}.
\subsection{ROUGE Score}
The quality of text summarization or machine translation output is assessed using a set of measures called ROUGE (Recall-Oriented Understudy for Gisting Evaluation). Comparing the generated text to the reference text forms the basis for the measurements. Precision, recall, and F1-score are used to construct ROUGE scores. The following is the ROUGE formula:

ROUGE-N:

$Precision = \frac{overlapping\ ngrams}{total\ ngrams}$\\
$Recall = \frac{number\ of\ overlapping\ ngrams} {number\ of\ ngrams\ in\ reference\ summary}$

$F1-score = 2* \frac{precision\ *\ recall}{precision\ +\ recall}$

The metrics reported in the paper are ROUGE-1 score. The score is calculated using the package rouge\_score.

\subsection{BLEU Score}
BLEU (Bilingual Evaluation Understudy) is used to assess the effectiveness by comparison of the generated text and the reference text forms the basis of it.

The nltk.translate.bleu\_score module in the NLTK package offers tools for computing BLEU scores. To compare a single generated sentence to a reference sentence and determine the BLEU score, use the sentence\_bleu() function. The sentence\_bleu() function allows you to specify the n-gram order (default is 4) and a set of weights to assign to each n-gram order. The weights are used to compute the final BLEU score, and they can be specified using the weights parameter. The weights parameter should be a tuple of floats that sum up to 1, where each float corresponds to the weight assigned to the n-gram order.

In this paper we have used sentence\_bleu with equal weigthage to all ngrams.

\subsection{Jaccard Similarity Score}
A measure of similarity between two sets of data is the Jaccard similarity score, commonly referred to as the Jaccard index or Jaccard coefficient. It is calculated by dividing the size of the intersection by the sum of the two sets. The following is the Jaccard similarity score formula:

$J(A, B) = \frac{|A \cap B|}{|A \cup B|}$

A and B are two sets, and the symbols for their intersection and union are and, respectively. The symbols |A| and |B| stand for the size or cardinality of the sets A and B, respectively.

The Jaccard similarity score is frequently used in text analysis to assess how similar two texts or text strings are to one another. The sets A and B can be defined as the set of words or tokens in the two documents, and the Jaccard similarity score can be used to measure the overlap between the sets of words.

\subsection{Accuracy}
Accuracy is used to check the correctness of the generated model. We calculated accuracy with the formulae: 
$Accuracy = \frac{total\ correct\ predictions}{total\ predictions}$

\section{Details on Guidelines}
\label{apx:guidelines}
This appendix is referred in section \ref{sec:cpgs}. For this paper, Acute Lymphoblastic Leukemia (ALL), Bone, and Kidney cancer types are used from NCCN guidelines to build DKG. ALL cancer guidelines is a 135-page document consisting of more than 35 pages of flowcharts and algorithms for decision-making, 59 pages of discussion, and the remaining pages for references to evidence. Bone cancer guidelines is a 102-page document consisting of 34 pages of flowcharts and algorithms for decision-making, 32 pages of discussion, and the remaining pages for references to evidence. Kidney cancer guidelines is an 81-page document consisting of 23 pages of flowcharts and algorithms for decision-making, 34 pages of discussion, and the remaining pages for references to evidence. 

\section{Clinical Practice Guideline Fragment}
\label{apx:cpgexample}
This Section is explanation of CPG Fragment shown in Figure \ref{fig:CPGFragmentFig}. The acronyms used are shown in Table \ref{tab:abbr}.

\begin{center}
\begin{table}[ht]
\begin{tabular}{ |c|c| } 
\hline
\textbf{Acronym} & \textbf{Abbreviation}\\
\hline
ALL & Acute Lymphoblastic Leukemia \\
Ph+ & Philadelphia chromosome-positive \\ 
AYA & Adolescents and Young Adults \\ 
TKI & Tyrosine Kinase Inhibitor \\ 
CR & Complete Recovery \\
MRD & Measurable Residual Disease \\
\hline
\end{tabular}
\caption{\label{tab:abbr} Shows the Abbreviations of Acronyms used in Figure \ref{fig:CPGFragmentFig}  }
\end{table}
\end{center}

This fragment shows how a Ph+ ALL patient should be treated. This Ph+ ALL is determined by the doctors after doing initial tests on the patient. Here AYA$^{n,p,q}$ tells that the panel believes patients in age range of 15–39 years. If the above patients do not have any substantial comorbidities. Substantial Comorbidities means having more than one illness at once. Then the patient is adviced with the following treatments Clinical trial or TKI+Chemotherapy$^{t}$ which means systematic chemotherpay or TKI+corticosteroid$^{t}$ which means systematic corticosteroid. After these treatments Response Assessment is done and classifies patient as either CR or less than CR. If he is observed as CR then MRD is monitored which can be persistent rising MRD or MRD- (MRD Negative). If the patient is of greater than 65 years of age then recommended treatments are Clinical trial or TKI+corticosteroid$^{t,cc}$ means we have to consider modifying dose according to patients age or TKI+chemotherapy$^{t,cc}$ i.e., dose modifications required. Then Response Assessment is done on the patient.


\end{document}